\documentclass[twocolumn,runningheads]{svjour2}
\smartqed  % flush right qed marks, e.g. at end of proof
\usepackage{graphicx}
\usepackage{mathptmx}      % use Times fonts if available on your TeX system
\journalname{Astrophysics and Space Science}
\begin{document}

\title{Associations of Very High Energy Gamma-Ray Sources Discovered by
       H.E.S.S. with Pulsar Wind Nebulae%\thanks{Grants or other notes
}

\titlerunning{Associations of H.E.S.S. VHE $\gamma$-ray Sources
              with Pulsar Wind Nebulae} % if too long for running head

\author{Yves A. Gallant        \and
        for the H.E.S.S. Collaboration %etc.
}


\institute{Y.A. Gallant \at
              Laboratoire de Physique Th\'eorique et Astroparticules \\
              UMR 5207, CNRS/IN2P3, Universit\'e Montpellier II \\
              34095 Montpellier Cedex 5, France \\
              Tel.: +33-4-6714 4189\\
              Fax:  +33-4-6714 4190\\
              \email{gallant@lpta.in2p3.fr}           %  \\
}

\date{Received: date / Accepted: date}

\maketitle

\begin{abstract}
The H.E.S.S. array of imaging Cherenkov telescopes has discovered
a number of previously unknown $\gamma$-ray sources in the very high
energy (VHE) domain above 100 GeV.  The good angular resolution
of H.E.S.S. ($\sim0.1^\circ$ per event), as well as its sensitivity
(a few percent of the Crab Nebula flux) and wide 5$^\circ$ field of
view, allow a much better constrained search for counterparts
in comparison to previous instruments.  In several cases, the
association of the VHE sources revealed by H.E.S.S. with pulsar
wind nebulae (PWNe) is supported by a combination of positional
and morphological evidence, multi-wavelength observations, and
plausible PWN model parameters.  These include the plerions in the
composite supernova remnants G\,0.9+0.1 and MSH~15--5{\it 2}, the
recently discovered Vela~X nebula, two new sources in the Kookaburra
complex, and the association of HESS\,J1825--137 with PSR\,B1823--13.
The properties of these better-established associations are
reviewed.  A number of other sources discovered by H.E.S.S. are
located near high spin-down power pulsars, but the evidence for
association is less complete.  These possible associations are
also discussed, in the context of the available multi-wavelength
data and plausible PWN scenarios.
\keywords{gamma rays: observations \and pulsars \and nebulae}
\PACS{98.70.Rz \and 97.60.Gb \and 98.38.-j}
\end{abstract}

% Your text comes here. Separate text sections with
% \section{Section title}
% \label{sec:1}
% and \cite{Ref1}
% \subsection{Subsection title}
% \label{sec:2}
% as required. Don't forget to give each section
% and subsection a unique label (see Sect.~\ref{sec:1}).
% \paragraph{Paragraph headings} Use paragraph headings as needed.
% \begin{equation}
% a^2+b^2=c^2
% \end{equation}

\section{Introduction}

   The High Energy Stereoscopic System (H.E.S.S.) is an array of
four imaging Cherenkov telescopes designed to study astrophysical
gamma-rays in the energy domain between about 100\,GeV and several
tens of TeV.  Its wide field of view and unprecedented sensitivity
in this energy range have allowed the discovery of a large number
of new very high energy (VHE) gamma-ray sources.  Several of these
are associated with pulsar wind nebulae (PWNe); they will be
individually reviewed in section \ref{established}, and their
general properties discussed.  Section \ref{candidates} will then
examine the criteria for establishing such PWN associations, and
discuss individual possible PWN counterparts for unidentified
H.E.S.S. sources.

\section{``Established'' VHE Pulsar Wind Nebulae}
\label{established}

\subsection{The Crab Nebula}

   The ``standard candle'' of very high energy (VHE) gamma-ray
astronomy will serve to introduce the emission mechanisms at play
in pulsar wind nebulae (PWNe).  The Crab Nebula is a bright source
of strongly polarised, non-thermal radiation across most of the
electromagnetic spectrum.  This emission, from the radio domain
up to high-energy gamma-rays below 1\,GeV, is generally interpreted
as synchrotron radiation from relativistic electrons and positrons
created and accelerated by the central pulsar.

   The higher-energy emission component observed in VHE gamma-rays,
and  by {\it EGRET} as unpulsed emission above 1\,GeV, is conventionally
interpreted as inverse Compton (IC) scattering by the same accelerated
electrons and positrons.  Target photons for the scattering process
include the cosmic microwave background (CMB), interstellar dust
and stellar emission, and at least in the case of the Crab, the
synchrotron photons themselves.  Hadronic emission models have also
been proposed for VHE emission from plerions (see \cite{Bed06,Horetc06});
in the present review, however, we will restrict ourselves to the
more conservative leptonic models consisting of synchrotron and IC
emission components.

   Observations of the Crab Nebula with H.E.S.S. have revealed
clear evidence for steepening at high energies of the VHE gamma-ray
spectrum, which can be described by a power law of photon index
$\Gamma = 2.39 \pm 0.03_{\rm stat}$ with an exponential cutoff
energy $E_{\rm c} = 14.3 \pm 2.1_{\rm stat}$\,TeV \cite{Crab}.
Such spectral curvature is consistent with expectations from
model calculations of the IC emission spectrum.

\subsection{VHE plerion in the composite SNR G\,0.9+0.1}

   The Crab Nebula is the prototype of a purely {\it plerionic}
supernova remnant (SNR), exhibiting a centre-filled morphology
and strongly polarised, non-thermal emission, properties which
are characteristic of PWNe.  The Galactic SNR G\,0.9+0.1 is of
the more general {\it composite} type, in which such a plerion
is found inside the shell of emission due to the supernova blast
wave \cite{HelBec87}. The composite morphology of G\,0.9+0.1 is
evident in the radio domain, and X-ray observations confirmed
the non-thermal nature of the plerionic emission component.  The
pulsar thought to power this PWN has however not been detected
up to now, presumably because its beaming is unfavourable.

   VHE gamma-ray emission from G\,0.9+0.1 was discovered
in deep H.E.S.S. observations of the Galactic Centre region
\cite{G0.9}.  The observed gamma-ray excess is well described
as a point source at a position consistent with that of the
plerion; given the precise point spread function of H.E.S.S.,
an upper limit of 1.3$'$ on the (assumed Gaussian) source
extension was derived.  This argues for the plerion rather
than the shell as the counterpart of the VHE source.
   The available radio, X-ray and VHE gamma-ray spectral data
are well described by a simple leptonic model, with a magnetic
field inside the plerion of 6\,$\mu$G, close to the equipartition
value. The dominant IC target photon component in this
model is from stellar photons rather than the CMB, as expected
for a source located in the central regions of the Galaxy.

\subsection{The nebula of PSR\,B1509--58 in MSH 15--5{\it 2}}

   The composite SNR MSH 15--5{\it 2}, also known as G\,320.4--1.2,
comprises an X-ray and radio-emitting shell which encloses a
bright, non-thermal X-ray nebula around the young pulsar
B1509--58 (see e.g.\ \cite{Gaeetc02}).  The good angular resolution
of H.E.S.S. allowed the discovery of an extended source of
VHE gamma-rays in this SNR, with a morphology similar to
that of the non-thermal X-ray nebula \cite{MSH15-52}.
This morphological correspondence, and the fact that the
available spectral data can be described by a simple
leptonic model with a plerion magnetic field value of
$\sim$17\,$\mu$G, motivate the identification of the PWN
as the source of the VHE emission.

   This H.E.S.S. discovery provided the first illustration of
the potential for VHE gamma-ray morphological studies of PWNe.
The synchrotron emission observed at lower frequencies
reflects the spatial distribution of a combination of the
accelerated electron density and the magnetic field strength,
and the latter can in general be quite non-uniform in PWNe.
By contrast, in a typical IC emission scenario the target
photons are approximately uniformly distributed on the
scale of the SNR, so that the VHE emission directly reflects
the spatial distribution of the high-energy electrons.
In the case of MSH 15--5{\it 2}, the observed VHE morphology
confirms that these electrons are predominantly distributed
along a NW-SE direction, which is thought to reflect the
rotation axis of the pulsar.

\subsection{The Vela X plerionic nebula}

   The Vela SNR is a large (diameter $\sim 8^\circ$), nearby
(distance $D \approx 290$\,pc) composite remnant.  It contains
a radio-emitting plerionic nebula, Vela X, powered by the young
and energetic Vela pulsar, PSR\,B0833--45.  Observations of
this region with H.E.S.S. revealed a very extended source of
VHE gamma-rays centered to the south of the pulsar
\cite{Vela-X}, overlapping a diffuse hard X-ray emission
feature first detected with {\it ROSAT} 
\cite{MarOge95} and aligned with a bright radio filament
within the plerion.

   In this PWN, the radio, X-ray and VHE gamma-ray
emission regions are all markedly offset from the pulsar
position.  This may be due to the supernova explosion occurring
in an inhomogeneous medium, and the resulting asymmetric reverse
shock displacing the PWN in the direction away from the higher
density medium \cite{BloCheFri01}.  Such an offset may be
typical of older PWNe; the Vela pulsar has a spin-down age
of 11\,kyr, significantly older than that of the Crab (1.2\,kyr)
or B1509--58 (1.7\,kyr). % G\,0.9+0.1 is also thought to be
% a few thousand years old.

   The VHE gamma-ray spectrum of this source significantly
steepens with increasing energy, and can be described by a
power law of photon index $\Gamma = 1.45 \pm 0.09_{\rm stat}
\pm 0.2_{\rm sys}$ with an exponential cutoff energy
$13.8 \pm 2.3_{\rm stat} \pm 4.1_{\rm sys}$\,TeV; this
constitutes the first clear measurement of a peak in the
spectral energy distribution at VHE energies \cite{Vela-X}.
Assuming the CMB is the main target photon
component for IC scattering in the outer regions of the Galaxy,
a total energy of $\sim 2 \times 10^{45}$\,erg in non-thermal
electrons between 5\,TeV and 100\,TeV could be deduced.
These results demonstrate how VHE observations of IC emission
allow direct inference of the {\em spatial} and {\em spectral}
distribution of non-thermal electrons in a PWN.

\subsection{Two new VHE sources in the Kookaburra}

   Although the distribution of target photons for IC emission
it not uniform in the Galaxy, it varies relatively smoothly in
contrast to the distribution of target material for hadronic
gamma-ray emission processes.  Moreover, the CMB provides a
minimum target photon density which is uniformly distributed.
Sensitive VHE gamma-ray observations should thus reveal any
sufficiently intense source of high-energy electrons in the
Galaxy, again in contrast to sources of high-energy hadrons,
for which the presence of dense target material is also a
necessary condition for detectability.  The survey of the
Galactic plane undertaken with H.E.S.S. thus has a strong
potential for detecting energetic PWNe.

   A survey of the Galactic plane performed with H.E.S.S. in 2005,
in the Galactic longitude range $300^\circ < \ell < 330^\circ$,
allowed the discovery of two new VHE sources located in the
Kookaburra complex of radio and X-ray emission
\cite{Kooka}.  The stronger of the two VHE sources,
HESS\,J1420--607, is most plausibly associated with the radio
and X-ray nebula of the energetic pulsar PSR\,J1420--6048.
The second source, HESS\,J1418--609, is similarly associated
with radio and X-ray emission exhibiting the properties of
PWN, the so-called Rabbit, though a pulsar has so far not been
clearly detected in this object \cite{NgRobRom05}.

   In both sources, the VHE emission has a large spatial extent
and is significantly offset from the pulsar position, which
may be due to ``crushing'' of the PWN by the SNR reverse shock
as hypothesised in the case of Vela~X, or perhaps to the effects
of rapid motion of the pulsar through the surrounding medium
\cite{Kooka}.  Both PWNe have been proposed
as possible counterparts of an unidentified {\it EGRET} source
coincident with the Kookaburra complex.  The clear separation
by H.E.S.S. of two VHE sources, coincident with each of the
two PWNe, illustrates the advantages of good angular resolution
in identifying the counterparts of gamma-ray sources.

\subsection{HESS\,J1825--137 as the nebula of PSR\,B1823--13}

   HESS\,J1825--137 is a strong VHE source discovered in the
first H.E.S.S. survey of the Galactic plane \cite{survey-I}.
PSR\,B1823--13, a pulsar with properties similar to that
of Vela, lies at its Northern edge, and exhibits an asymmetric
X-ray nebula extending in the direction of the centre of the
VHE source \cite{Gaeetc03}.
The detected diffuse X-ray emission only extends over $\sim$5$'$,
however, much smaller than the size of the VHE source.  In
contrast to the previously discussed sources, there is no
good morphological match of HESS\,J1825--137 with emission
detected at other wavelengths.

   Morphological studies of HESS\,J1825--137 in the VHE gamma-ray
domain have nonetheless yielded compelling evidence for its association
with PSR\,B1823--13.  In particular, the VHE emission has an
asymmetric profile with a sharp peak immediately South of the
pulsar position; the shape of this profile is similar to that
of the X-ray nebula, but the VHE profile extends over a much
larger scale \cite{J1825-I}.
More importantly, deeper H.E.S.S. observations have revealed
the energy-dependent morphology of HESS\,J1825--137, marking
the first time such an effect is detected in VHE gamma-rays.
This manifests itself as a steepening of the power-law spectral
index with increasing distance from PSR\,B1823--13, as would be
expected from radiative losses of high-energy electrons injected
by the pulsar \cite{J1825-II}.  These losses could also account
for the fact that the PWN appears larger in VHE gamma-rays
than in X-rays, as in a leptonic scenario the latter are
emitted by higher-energy electrons.

\section{VHE Pulsar Wind Nebula Candidates}
\label{candidates}

\subsection{Association criteria}

   A total of seven fairly well-established associations of
VHE gamma-ray sources with PWNe have been reviewed in section
\ref{established}.  To these might be added the VHE emission
associated with PSR\,B1259--63 \cite{B1259}.
This object is in a different class from the other PWNe
and candidate PWNe discussed above and below, however,
in that it is dominated by the interaction with its binary
companion, as evidenced by its orbital variability;
it will thus not be considered further here.

   For six of the VHE sources discussed in section \ref{established},
the association rests on a positional and morphological match to
a PWN known at lower energies.  When this is not the case, i.e.\
for HESS\,J1825--137, an alternative criterion is morphological
and spectral evidence in the VHE gamma-ray domain for association
with a known pulsar, and consistent data at other wavelengths.
In all cases, another necessary criterion is a physically
plausible spectral model which is consistent with the available
multi-wavelength (MWL) data on the object.
In this section we will examine other candidate associations, for
which the above criteria are not currently all fulfilled with the
available MWL and VHE data.

\subsection{Pulsar energetics}

   When considering the association of a VHE source with the
nebula of a known pulsar, an additional criterion is the apparent
efficiency for VHE gamma-ray emission, given the pulsar's current
spin-down luminosity $\dot{E}$.  The comparison assumes that the
VHE source is located at the pulsar distance $D$, generally
determined from the radio dispersion measure.  For definiteness,
we use the VHE energy flux $F_{0.3-30}$ integrated over the
the energy range 0.3--30\,TeV.  This is roughly representative of
the H.E.S.S. spectral analysis range, although the energy threshold
for individual sources depends on the observation zenith angle,
and the upper limit depends on photon statistics.
The apparent efficiency $\varepsilon$ is then defined as $
% \begin{equation}
\varepsilon \equiv \left( 4 \pi D^2 F_{0.3-30} \right) / \dot{E} $.
% \; . \end{equation}

%\begin{table}[t]
%\centering
%\begin{tabular}{lll}
%\hline\noalign{\smallskip}
%first & second & third  \\[3pt]
%\tableheadseprule\noalign{\smallskip}
%umber & number & number \\
%\end{tabular}
%\end{table}

\begin{table}[t]
\caption{Apparent efficiencies for ``established'' associations}
\centering
\label{eff_estab}
\begin{tabular}{llll}
\hline\noalign{\smallskip}
VHE source & $F_{0.3-30}$ $^{\rm a}$ & PSR name & $\varepsilon$ \\[3pt]
\tableheadseprule\noalign{\smallskip}
Crab nebula      & $1.7 \times 10^{-10}$     & B0531+21    & 0.02\% \\
MSH 15--5{\it 2} & $3.3 \times 10^{-11}$     & B1509--58   & 0.4\% \\
Vela X           & $9\;\;\;\: \times 10^{-11}$ & B0833--45 & 0.01\% \\
HESS\,J1420--607 & $2.2 \times 10^{-11}$     & J1420--6048 & 0.8\% \\
HESS\,J1825--137 & $1.1 \times 10^{-10}$     & B1823--13   & 7\% \\
\noalign{\smallskip}\hline
\end{tabular}
\flushleft \hspace{6mm}
$^{\rm a}$ in units of erg\,cm$^{-2}$\,s$^{-1}$
\end{table}

   Table \ref{eff_estab} list the VHE energy fluxes and apparent
efficiencies for the five well-established VHE PWNe in which the
pulsar has been detected and timed.  The fluxes $F_{0.3-30}$ were
obtained by integration of the best-fit spectral model as given
in the references listed in section \ref{established}.  The pulsar
parameters $\dot{E}$ and $D$ were obtained from the ATNF Pulsar
Catalogue \cite{Manetc05}, version 1.25,
using the NE2001 model of the Galactic free electron distribution
for the distance \cite{CorLaz02}.

   The apparent efficiency reflects the true efficiency only
to the extent that the emitting particles' lifetimes are short
compared with the evolutionary time scale of the PWN.  In general
the VHE-emitting electrons may have been injected in the early
phases of the PWN evolution, when the pulsar's $\dot{E}$ was
larger, so that the apparent efficiency is an overestimate of
the true efficiency; this appears to be the case in particular
for HESS\,J1825--137 \cite{J1825-II}.
Nonetheless, associations for which the required efficiency
approaches 100\% may be considered questionable, and those
for which it far exceeds this can generally be ruled out as
implausible.

\subsection{Possible associations with known pulsars}

   In addition to HESS\,J1825--137, two other VHE sources
discovered in the initial H.E.S.S. survey of the Galactic
plane may be associated with energetic pulsars
\cite{survey-I}.  Located near the edge of the bright
source HESS\,J1616--508 is PSR\,J1617--5055, an X-ray
emitting young pulsar with a period of 69\,ms and a
spin-down luminosity $\dot{E} = 1.6 \times 10^{37}$\,erg/s.
Although its association with the VHE source is energetically
plausible, the putative wind nebula of this pulsar has not
been detected at other wavelengths.

   One of the brightest and largest sources discovered in
the Galactic plane survey, HESS\,J1804--216, contains the
young and energetic pulsar B1800--21, with spin-down
luminosity $\dot{E} = 2.2 \times 10^{36}$\,erg/s.  As in
the previous case, an association is energetically plausible,
but no coincident PWN has been detected at other wavelengths.
Alternatively, the H.E.S.S. source could be associated with
part of the shell-type SNR G\,8.7--0.1 \cite{survey-II}.

   Although HESS\,J1303--631, the first unidentified source
discovered by H.E.S.S., has no established counterpart, it
does coincide with the energetic pulsar J1301--6305, with
spin-down luminosity $\dot{E} = 1.7 \times 10^{36}$\,erg/s.
Its catalogued distance using the Galactic free electron
model of Taylor and Cordes \cite{TayCor93} was $D = 15.8$\,kpc,
which required a very high efficiency of order 40\% to
power the VHE source \cite{J1303}.  The more recent NE2001
model, however, implies a distance of only $D = 6.65$\,kpc,
making the apparent efficiency comparable with that for
HESS\,J1825--137 (see Table \ref{eff_candid}).

   One additional possible association is with HESS\,J1702--420
discovered in the Galactic plane survey.   The nearby pulsar
J1702--4128 would require a high but not impossible efficiency
to power the entire H.E.S.S source.  It is located near the
tip of a tail-like extension from HESS\,J1702--420.  Although
this tail was not statistically significant in the original
survey data \cite{survey-II}, additional
H.E.S.S. observations have since increased its significance.
The offset of PSR\,J1702--4128 from the core of HESS\,J1702--420
is large, making an association less likely, but it may be that
only part of the H.E.S.S. source is associated with the
nebula of PSR\,J1702--4182.  If such an association were
confirmed, with a spin-down age of 55\,kyr and luminosity
$\dot{E} = 3.4 \times 10^{35}$\,erg/s this would be the oldest
and least energetic pulsar yet found to have a VHE-emitting
wind nebula.

\begin{table}[t]
\caption{Required efficiencies for candidate associations}
\centering
\label{eff_candid}
\begin{tabular}{llll}
\hline\noalign{\smallskip}
VHE source & $F_{0.3-30}$ $^{\rm a}$ & PSR name & \ $\varepsilon$ \\[3pt]
\tableheadseprule\noalign{\smallskip}
HESS\,J1616--508 & $3.7 \times 10^{-11}$ & J1617--5055 & \ 1.3\% \\
HESS\,J1804--216 & $2.9 \times 10^{-11}$ & B1800--21   & \ 2.4\% \\
HESS\,J1303--631 & $2.3 \times 10^{-11}$ & J1301--6305 & \ 7\% \\
HESS\,J1702--420 & $1.4 \times 10^{-11}$ & J1702--4128 & 11\% \\
\noalign{\smallskip}\hline
\end{tabular}
\flushleft \hspace{6mm}
$^{\rm a}$ in units of erg\,cm$^{-2}$\,s$^{-1}$
\end{table}

   Table \ref{eff_candid} summarises the VHE energy fluxes and
required efficiencies for these candidate associations; the numbers
were derived in the same manner as in Table~\ref{eff_estab}.
In all cases the pulsar is significantly offset from the centre
of the VHE source, but as was seen in section \ref{established},
this would seem to be typical of older VHE PWNe.  Deeper MWL or
VHE observations would be necessary in order to establish any of
these candidate associations.

\subsection{Possible VHE PWNe without detected pulsars}

   The example of G\,0.9+0.1 shows that H.E.S.S.-discovered sources
can be associated with PWNe even if the corresponding pulsar has
not been detected, in particular when the VHE source is coincident
with a composite SNR.  Another possible such association is with
HESS\,J1813--178; this relatively compact VHE source was
discovered in the Galactic plane survey \cite{survey-I}, and was
subsequently found to be coincident with a shell-type radio SNR,
G\,12.82--0.02 \cite{Broetc05,HelBecWhi05}, and a bright, non-thermal,
hard X-ray source \cite{Broetc05,Ubeetc05}.
The angular resolution of H.E.S.S. or of the available X-ray data
could however not discriminate between the shell and a possible
embedded PWN as the source of the respective emission.  A recent
{\it XMM-Newton} observation of this region shows evidence for
a PWN origin of the X-ray emission, suggesting a composite nature
for G\,12.82--0.02 and the possibility of a PWN origin for the
VHE emission \cite{Fun06}.

   Another source discovered in the Galactic plane survey is
HESS\,J1834--087, which is positionally coincident with the radio
SNR G\,23.3--0.3, also known as W41.  The VHE source extension
appears to be smaller than the radius of the shell, and its
position coincides with a region of enhanced radio emission near
the centre of the shell \cite{survey-II}.  This suggests the
intriguing possibility that W41 might be a composite SNR, and
the VHE emission might originate in a central plerion; more MWL
observations of this SNR are needed to support such a scenario,
however.  An alternative possibility is that the VHE emission
is due to hadronic processes and originates in a large molecular
cloud associated with W41, which is in good positional coincidence
with the VHE source \cite{Albetc06}.

   As a final example, one of the potential counterparts suggested
for the Galactic survey source HESS\,J1634--472 \cite{survey-II}
was the radio SNR candidate G\,337.2+0.1, coincident with an X-ray
source detected by {\it ASCA} \cite{Cometc05}.  A recent {\it
XMM-Newton} observation of this region shows evidence for a PWN
origin of the X-ray emission \cite{Cometc06} and
raises the possibility of a PWN association for the VHE emission.
The relatively small angular size of this candidate radio and X-ray
PWN compared with that of HESS\,J1634--472, and its location near
the edge of the VHE excess, nonetheless make an association with
the whole of the VHE source unlikely.

\section{Summary and Prospects}

   Of the VHE gamma-ray sources detected by H.E.S.S., seven have
fairly well-established PWN counterparts, not including the VHE
emission associated with PSR\,B1259--63.  These currently constitute
the most numerous class of identified Galactic VHE gamma-ray
sources.  Several of these VHE-emitting PWNe exhibit a large physical
extent and are significantly offset from the pulsar position; one
possible explanation is that these are older PWNe, strongly affected
by the passage of an asymmetric reverse shock in the parent SNR.

   In a leptonic interpretation of the VHE emission, the target
photons for IC scattering have an approximately known and uniform
density in individual PWNe, which allows direct inference of the
{\em spectral} and {\em spatial} distribution of the energetic
electrons, in contrast to observations of syn\-chro\-tron emission
at lower energies.  VHE gamma-ray astronomy thus provides a new,
independent observational window into the physics of PWNe.

   Given smoothly varying Galactic target photon densities, and the
uniform target density provided by the Cosmic Microwave Background,
a survey in VHE gamma-rays should reveal all sufficiently intense
Galactic sources of high-energy electrons.  Four more VHE
sources discovered by H.E.S.S. may be associated with known
energetic pulsars, and three additional such sources are coincident
with possible PWNe in which the pulsar has not been detected.
More observations of these sources in VHE gamma-rays and at
other wavelengths are necessary to investigate the possibility
of these associations.  PWNe may yet prove to constitute the
fastest-growing class of identified Galactic gamma-ray sources.

\begin{acknowledgements}
The support of the Namibian authorities and of the University of
Namibia in facilitating the construction and operation of H.E.S.S.
is gratefully acknowledged, as is the support by the German Ministry
for Education and Research (BMBF), the Max Planck Society, the
French Ministry for Research, the CNRS-IN2P3 and the Astroparticle
Interdisciplinary Programme of the CNRS, the U.K. Particle Physics
and Astronomy Research Council\break (PPARC), the IPNP of the Charles
University, the South African Department of Science and Technology
and National Research Foundation, and by the University of Namibia.
We appreciate the excellent work of the technical support staff in
Berlin, Durham, Hamburg, Heidelberg, Palaiseau, Paris, Saclay and
in Namibia in the construction and operation of the equipment.
\end{acknowledgements}



\begin{thebibliography}{3}
\bibitem{Bed06}
Bednarek, W.:
High Energy Processes in Pulsar Wind Nebulae.
In: these proceedings.
{\tt astro-ph/0610307} (2006)

\bibitem{Horetc06}
Horns, D., Aharonian, F., Hoffmann, A.I.D., Santangelo, A.:
Nucleonic gamma-ray production in Pulsar Wind Nebulae.
In: these proceedings.
{\tt astro-ph/0609386} (2006)

\bibitem{Crab}
Aharonian, F., et al.\ ({\it H.E.S.S. Collaboration}):
Observations of the Crab nebula with HESS.
A\&A {\bf 457}, 899--915 (2006)

\bibitem{HelBec87}
Helfand, D.J., Becker, R.H.:
G\,0.9+0.1 and the emerging class of composite supernova remnants.
ApJ {\bf 314}, 203--214 (1987)

\bibitem{G0.9}
Aharonian, F., et al.\ ({\it H.E.S.S. Collaboration}):
Very high energy gamma rays from the composite SNR G\,0.9+0.1.
A\&A {\bf 432}, L25--L29 (2005)

\bibitem{Gaeetc02}
Gaensler, B.M., Arons, J., Kaspi, V.M., Pivovaroff, M.J., Kawai, N.,
Tamura, K.:
Chandra Imaging of the X-Ray Nebula Powered by Pulsar B1509--59.
ApJ {\bf 569}, 878--893 (2002)

\bibitem{MSH15-52}
Aharonian, F., et al.\ ({\it H.E.S.S. Collaboration}):
Discovery of extended VHE gamma-ray emission from the asymmetric pulsar
wind nebula in MSH 15--5{\it 2} with HESS.
A\&A {\bf 435}, L17--L20 (2005)

\bibitem{Vela-X}
Aharonian, F., et al.\ ({\it H.E.S.S. Collaboration}):
First detection of a VHE gamma-ray spectral maximum from a cosmic source:
HESS discovery of the Vela X nebula.
A\&A {\bf 448}, L43--L47 (2006)

\bibitem{MarOge95}
Markwardt, C.B., \"Ogelman, H.:
An X-Ray jet from the Vela Pulsar.
Nature {\bf 375}, 40--42 (1995)

\bibitem{BloCheFri01}
Blondin, J.M., Chevalier, R.A., Frierson, D.M.:
Pulsar Wind Nebulae in Evolved Supernova Remnants.
ApJ {\bf 563}, 806--815 (2001)

\bibitem{Kooka}
Aharonian, F., et al.\ ({\it H.E.S.S. Collaboration}):
Discovery of the two ``wings'' of the Kookaburra complex in VHE
$\gamma$-rays with HESS.
A\&A {\bf 456}, 245--251 (2006)

\bibitem{NgRobRom05}
Ng, C.-Y., Roberts, M.S.E., Romani, R.W.:
Two Pulsar Wind Nebulae: Chandra/XMM-Newton Imaging of GeV J1417-6100.
ApJ {\bf 627}, 904--909 (2005)

\bibitem{survey-I}
Aharonian, F., et al.\ ({\it H.E.S.S. Collaboration}):
A New Population of Very High Energy Gamma-Ray Sources in the Milky Way.
Science {\bf 307}, 1938--1942 (2005)

\bibitem{Gaeetc03}
Gaensler, B.M., Schulz, N.S., Kaspi, V.M., Pivovaroff, M.J., Becker, W.E.:
XMM-Newton Observations of PSR B1823--13: An Asymmetric Synchrotron Nebula
around a Vela-like Pulsar.
ApJ {\bf 588}, 441--451 (2003)

\bibitem{J1825-I}
Aharonian, F.A., et al.\ ({\it H.E.S.S. Collaboration}):
A possible association of the new VHE $\gamma$-ray source HESS J1825--137
with the pulsar wind nebula G\,18.0--0.7.
A\&A {\bf 442}, L25--L29 (2005)

\bibitem{J1825-II}
Aharonian, F., et al.\ ({\it H.E.S.S. Collaboration}):
Energy-dependent gamma-ray morphology in the pulsar wind nebula
HESS J1825--137.
A\&A (in press). {\tt astro-ph/0607548} (2006)

\bibitem{B1259}
Aharonian, F. et al.:
Discovery of the binary pulsar PSR B1259--63 in very-high-energy
gamma rays around periastron with HESS.
A\&A {\bf 442}, 1--10 (2005)

\bibitem{Manetc05}
Manchester, R.N., Hobbs, G.B., Teoh, A., Hobbs, M.:
The Australia Telescope National Facility Pulsar Catalogue.
AJ {\bf 129}, 1993--2006 (2005),
%{\tt
http://www.atnf.csiro.au/research/pulsar/psrcat/
%}

\bibitem{CorLaz02}
Cordes, J.M., Lazio, T.J.W.:
NE2001. I. A New Model for the Galactic Distribution of Free Electrons
and its Fluctuations.
{\tt astro-ph/0207156} (2002)

\bibitem{survey-II}
Aharonian, F., et al.\ ({\it H.E.S.S. Collaboration}):
The H.E.S.S. Survey of the Inner Galaxy in Very High Energy Gamma Rays.
ApJ {\bf 636}, 777--797 (2006)

\bibitem{TayCor93}
Taylor, J.H., Cordes, J.M.:
Pulsar distances and the Galactic distribution of free electrons.
ApJ {\bf 411}, 674--684 (1993)

\bibitem{J1303}
Aharonian, F., et al.\ ({\it H.E.S.S. Collaboration}):
Serendipitous discovery of the unidentified extended TeV $\gamma$-ray
source HESS J1303--631.
A\&A {\bf 439}, 1013--1021 (2005)

\bibitem{Broetc05}
Brogan, C.L., Gaensler, B.M., Gelfand, J.D., Lazendic, J.S., Lazio,
T.J.W., Kassim, N.E., McClure-Griffiths, N.M.:
Discovery of a Radio Supernova Remnant and Nonthermal X-Rays Coincident
with the TeV Source HESS J1813--178.
ApJ {\bf 629}, L105--L108 (2005)

\bibitem{HelBecWhi05}
Helfand, D.J., Becker, R.H., White, R.L.:
A Radio Counterpart for the Unidentified TeV Source HESS J1813-178:
The Radio-Gamma-Ray Connection.
{\tt astro-ph/0505392} (2005)

\bibitem{Ubeetc05}
Ubertini, P., Bassani, L., Malizia, A., Bazzano, A., Bird, A.J.,
Dean, A.J., De Rosa, A., Lebrun, F., Moran, L., Renaud, M., Stephen,
J.B., Terrier, R., Walter, R.:
INTEGRAL IGR J18135-1751 = HESS J1813-178: A New Cosmic High-Energy
Accelerator from keV to TeV Energies.
ApJ {\bf 629}, L109--L112 (2005)

\bibitem{Fun06}
Funk, S.: %, for the H.E.S.S. Collaboration:
Status of Identification of VHE $\gamma$-ray sources.
In: these proceedings.
{\tt astro-ph/0609586} (2006)

% \bibitem{Whietc06}
% Helfand, D.J., Becker, R.H., White, R.L., Fallon, A., Tuttle, S.:
% MAGPIS: A Multi-Array Galactic Plane Imaging Survey.
% AJ {\bf 131}, 2525--2537 (2006)

\bibitem{Albetc06}
Albert, J., et al.\ ({\it MAGIC Collaboration}):
Observation of VHE Gamma Radiation from HESS J1834--087/W41 with the
MAGIC Telescope.
ApJ {\bf 643}, L53--L56 (2006)

\bibitem{Cometc05}
Combi, J.A., Benaglia, P., Romero, G.E., Sugizaki, M.:
G337.2+0.1: A new X-ray supernova remnant?
A\&A {\bf 431}, L9--L12 (2005)

\bibitem{Cometc06}
Combi, J.A., et al.:
XMM detection of the SNR G\,337.2+0.1.
In: these proceedings

% \bibitem{Ref1}
%Author, I.: Article title. Journal Title-Abbreviated {\bf Vol}, pp--pp (year)
%Format for books
%\bibitem{Ref2}
%Author, I., Smith, J.: Book Title. Publisher, Place (year)
% Format for proceedings
%\bibitem{Ref3}
%Author, I., Smith, J.: Paper title. In: Editor, A. (ed.) Proceedings
%Title, Location, Date, pages. Publisher, Place (year)
% etc
\end{thebibliography}
\end{document}